# In the Aftermath of Oil Prices Fall of 2014/2015–Socioeconomic Facts and Changes in the Public Policies in the Sultanate of Oman

*Osama A. Marzouk*

University of Buraimi, College of Engineering
P.O. Box 890, P.C. 512 Al Buraimi, Sultanate of Oman

**Abstract:** Since the start of its national renaissance in 1970, the Sultanate of Oman (Oman) has gone over a major development in several areas, such as education, infrastructure, and urbanization. This has been powered by the revenues from exporting crude oil and natural gas, which together form the skeleton of the country's economy. In the second half of 2014, the oil prices declined strongly to about 50% of its price. This was followed by another moderate decline in the second half of 2015 and the beginning of 2016, leaving the barrel price at a low level below 30 US$ in January 2016 (as compared to above 110 US$ in June 2014). This drop had direct impacts on the economy of Oman, manifested in a large budget deficit, reduced governmental expenditure, reduced or cancelled subsidy of fuels and electricity, increase in the water tariff, and decline in deposits in banks. The country is coping with this through its 9[th] five-year plan (2016-2020), which adopts a strategy of diversifying the income and relying less on the traditional oil and gas sector. The country has also taken measures to facilitate private businesses. This article sheds light on these topics as well as miscellaneous data about Oman.

**Keywords:** Oman; Oil Price Fall; Policies; Budget; Diversification

## 1. About Oman

The Sultanate of Oman (Oman) is a medium-sized Arab country that lies in the south-eastern tip of the Arabian Peninsula. Its area is 309,500 km² **[1]**. The neighbor countries of Oman are the United Arab Emirates (UAE) in the northwest, Saudi Arabia in the west, and Yemen in the southwest. It has a long coastline of about 3,165 km, covering three bodies of water: the Arabian Gulf (which is also known as the Persian Gulf), the Gulf of Oman, and the Arabian Sea. The capital is Muscat, and it is located in the northeast coast of Oman, on the Gulf of Oman. Figure 1 shows two maps that demonstrate the location of Oman.

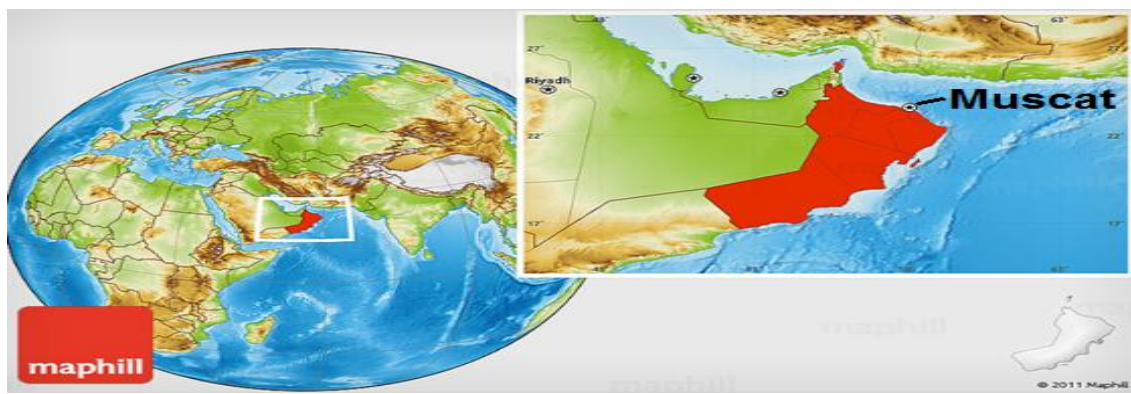

**Figure 1:** A physical location map of Oman [2] and its Capital. Adapted from the original map, which is available freely.

The estimated population of Oman (regardless of legal status or citizenship) in 2015 was 4,490,500 [**3**]. As of March 2016, about 45% of this population was expatriates, who reached 2 million according to the National Centre for Statistics and Information (NCSI). Most (about 1,747,000) of these expatriates are working personnel with no dependents. The majority (about 75%) of these expatriates are Indians (about 670,000 or 34%); Bangladeshis (about 590,000 or 30%); and Pakistanis (about 220,000 or 11%). The citizens (Omanis) in Muscat represent 36.2% of its population [**4**]. These figures show how diverse the country is.

Figure 2 shows the historical increase in the country's population (from 1960 to 2015). The total population was 551,737 in 1960. This reached 4,490,541 in 2015. Thus, the population increased more than 8 times in 55 years. The increase has been steepest since 2009 (2,762,073). Between 2009 and 2015, the population increased by a factor of 1.63.

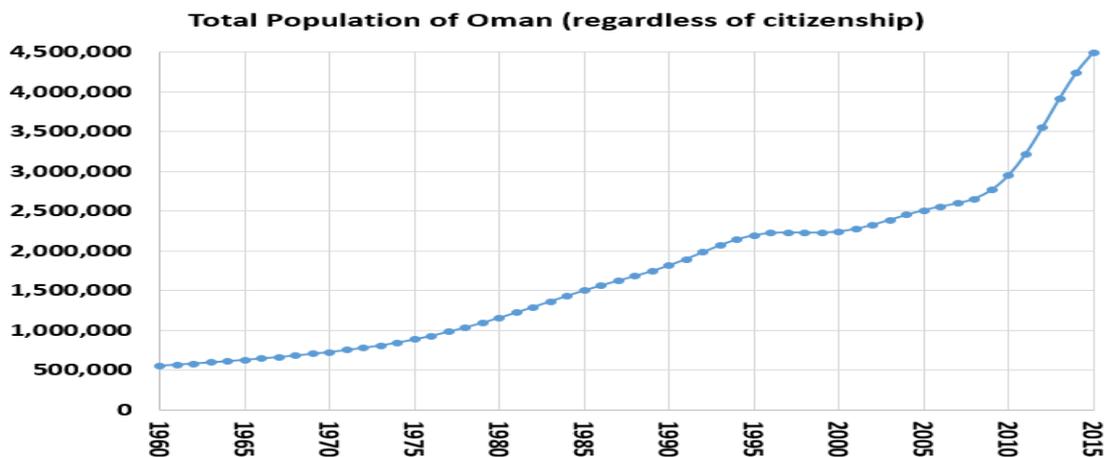

**Figure 2:** Historical change in the total population of Oman. Data taken from [3].

For a better realization of the population change, the growth rate is given in Figure 3. The growth rate is defined here as the exponential rate of growth of midyear population from year (t-1) to (t), expressed as a percentage. It does not account for the legal status or citizenship, but just the presence inside the country.

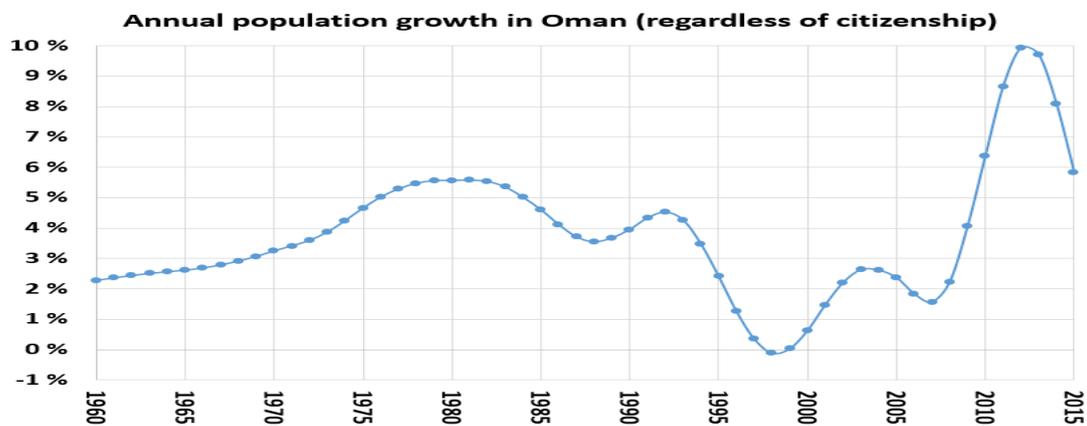

**Figure 3**: Historical change in the exponential growth rate of the total population of Oman. Data taken from [5].

The local currency is the Omani rial (OMR), and it is pegged to U.S. dollar **[6]** at 1 OMR = 2.6 US$, and this has been a steady exchange rate since 1986. The Omani rial is the third-highest-valued currency unit, coming after the Kuwaiti dinar and the Bahraini dinar. Regarding the direct foreign investments into Oman, 46% of them are from the U.K. **[7]**.

The government system in Oman is royal, and the head of the government is His Majesty Sultan Qaboos bin Said, who enjoys internal and external respect. His wise foreign policies have maintained peaceful relations between Oman and other countries. The 2016 Global Peace Index (GPI) for Oman is 2.016 (middle band), and it is ranked 74[th] (out of 163). This index considers 23 qualitative and quantitative indicators. A smaller score corresponds to more peace **[8]**. This ranking makes Oman much more peaceful than its neighbor Yemen (GPI = 3.399, ranked 158/163), more peaceful than its neighbor Saudi Arabia (GPI = 2.338, ranked 129/163), but slightly less peaceful than its other neighbor UAE (GPI = 1.931, ranked 61/163).

Oman has a long and interesting history, going back thousands of years to the times of the civilization of Magan [9]. Oman was once a powerful empire after overcoming the Portuguese forces in the 17[th] century, extending from India in the east to Zanzibar (East Africa) in the west, with the capital being Muscat as it is today [10]. There are no rivers, and the country's sources of water include groundwater, rainfalls, and water desalination. The country-wise average precipitation is 125 mm/year [11]. However, this varies largely from one zone to another.

## 2. Development in Oman

Since the national renaissance in 1970, Oman has developed remarkably in several fields. This is manifested in stronger infrastructure, more schools and universities, broader health sector, and industrialization. Figure4 shows the historical data of the gross domestic product (GDP) between 1965 and 2015. The sharp drop in 2009 and 2015 is attributed to sharp declines in the oil price. Nevertheless, there has been a monotonic steep increase between 2004 and 2008, and between 2009 and 2012. The growth in the GDP in these periods is far steeper than how it was in the 1990s and early 2000s.

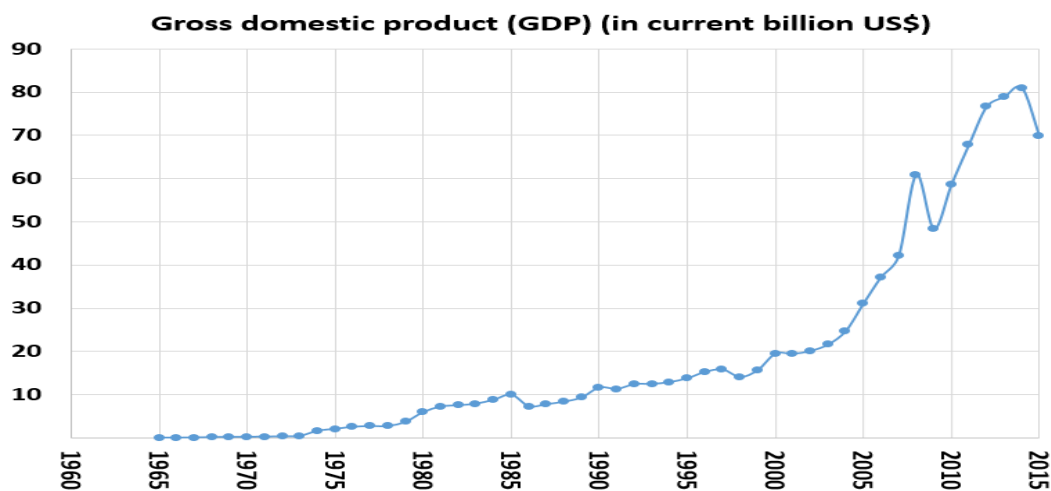

**Figure 4:** Historical change in the gross domestic product for Oman (converted to billion U.S. dollars). Data taken from [12].

Figure 5 shows similar data for the gross national income (GNI) per capita between 1967 and 2015. The gross national income per capita also increased rapidly between 2003 and 2008. After that, it slowed down. Like the GDP, it dropped following the oil price falls in 2014, and the continuous increase in population not matched with an increase in the income.

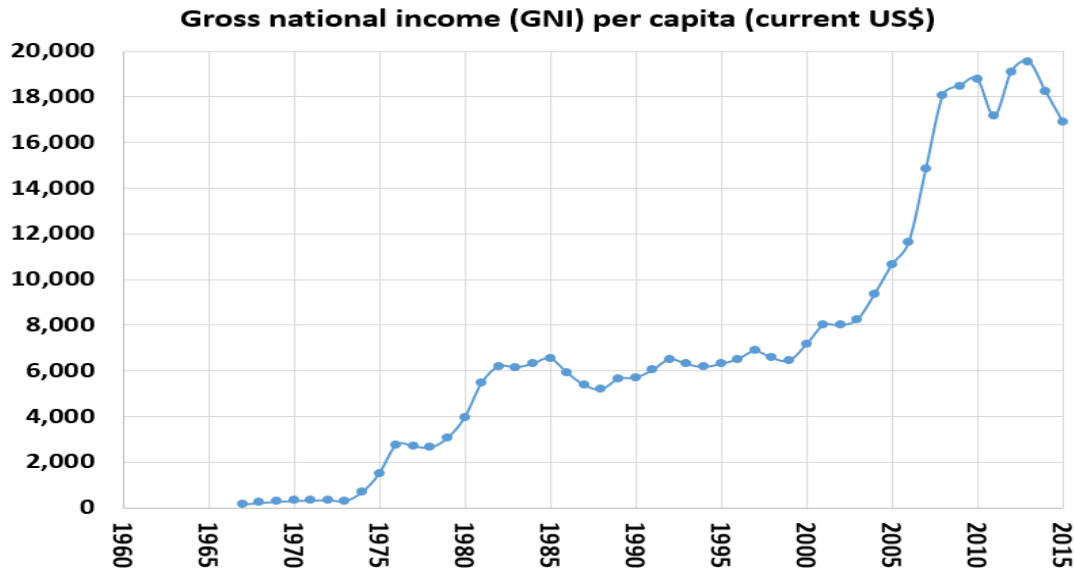

**Figure 5:** Historical change in the gross national income per capita for Oman (converted to U.S. dollars). Data taken from [13].

As an evidence for the development in the health care in Oman, the historical data of the life expectancy is given in Figure 6. The life expectancy increased from 42.7 years in 1960 to 77.1 years in 2014. This reflects remarkable reductions in the infant mortality rate (defined as the infant deaths per 1,000 live births). Table 1 shows how the infant mortality rate in Oman decreased since 1990 and how it is expected to decrease further through 2100 [14].

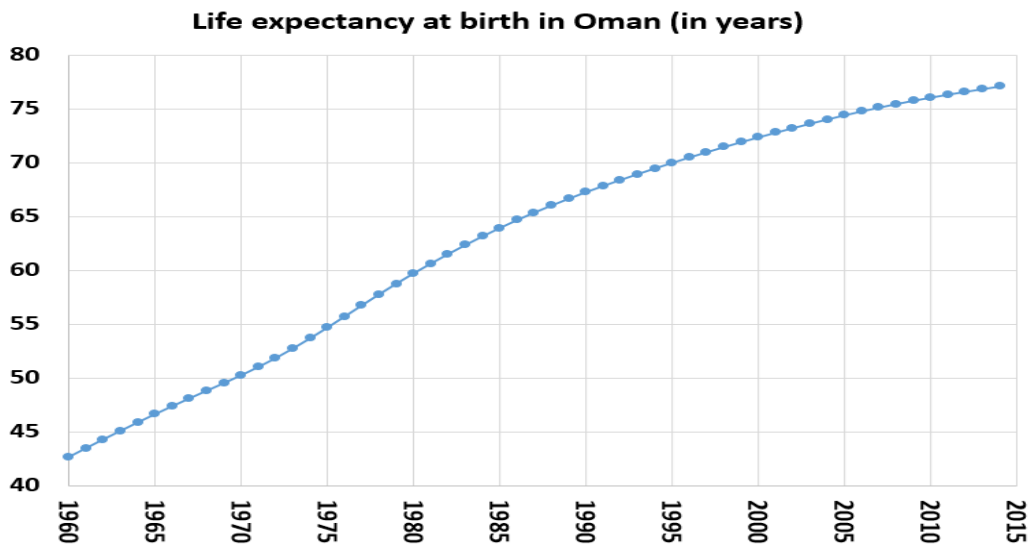

**Figure 6:** Historical change in life expectancy in Oman (in years). Data taken from[15].

**Table 1:** Infant mortality rate in Oman (past data and forecast)

| Year | 1990-1995 | 2005-2010 | 2010-2015 | 2015-2020 | 2025-2030 | 2045-2050 | 2095-2100 |
|------|-----------|-----------|-----------|-----------|-----------|-----------|-----------|
| | 30.9 | 9.8 | 7.3 | 5.9 | 4.6 | 3.2 | 1.8 |

Oman ranked 52nd in the 2015 Human Development Index (HDI) rankings [16]. It was (with Belarus) the 1st among the high-human-development category. The HDI score of Oman was 0.796. Table 2 demonstrates the trend of the HDI for Oman.

**Table 2:** Trend of the Human Development Index (HDI) for Oman

| Year | 2000 | 2010 | 2011 | 2012 | 2013 | 2014 | 2015 |
|------|------|------|------|------|------|------|------|
| | 0.705 | 0.797 | 0.797 | 0.796 | 0.796 | 0.795 | 0.796 |

In terms of urbanization, it is progressing rapidly in Oman, compared to other countries, as shown in Figure 7. The vertical line in the figure corresponds to the average annual growth rate of the urban population of Oman between 1950 and 2014. The gray area around that vertical line represents the average annual growth rates of the urban population of all countries of the world. The average growth rate in Oman is about 3.4%, and puts Oman in a leading position.

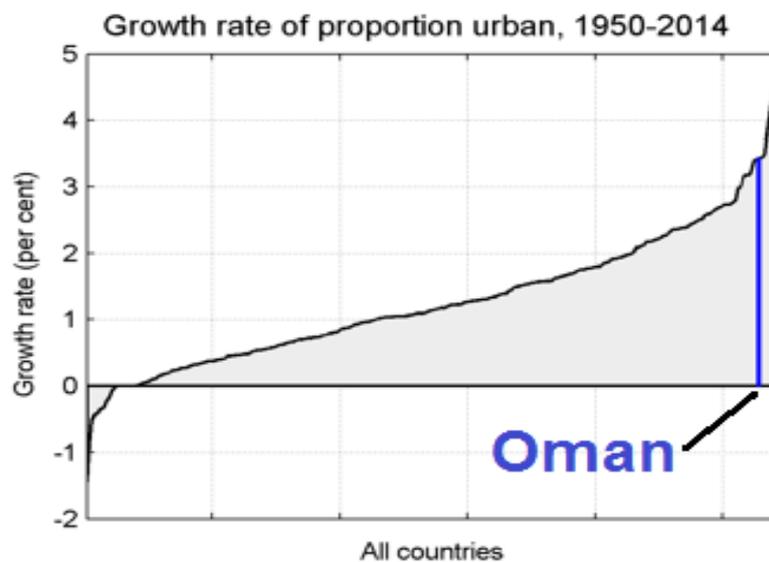

**Figure 7:** Average growth rate of proportion urban for Oman as compared to other countries. Figure adapted from [17].

Like neighbor countries, Oman has been depending chiefly on its oil and gas for the revenues needed to continue its development plans. Figure 8 shows the production and exports of crude oil in Oman between the beginning of 2004 and the beginning of 2016. The gap between them is not large, reflecting that the country exports the majority of its production in its crude form.

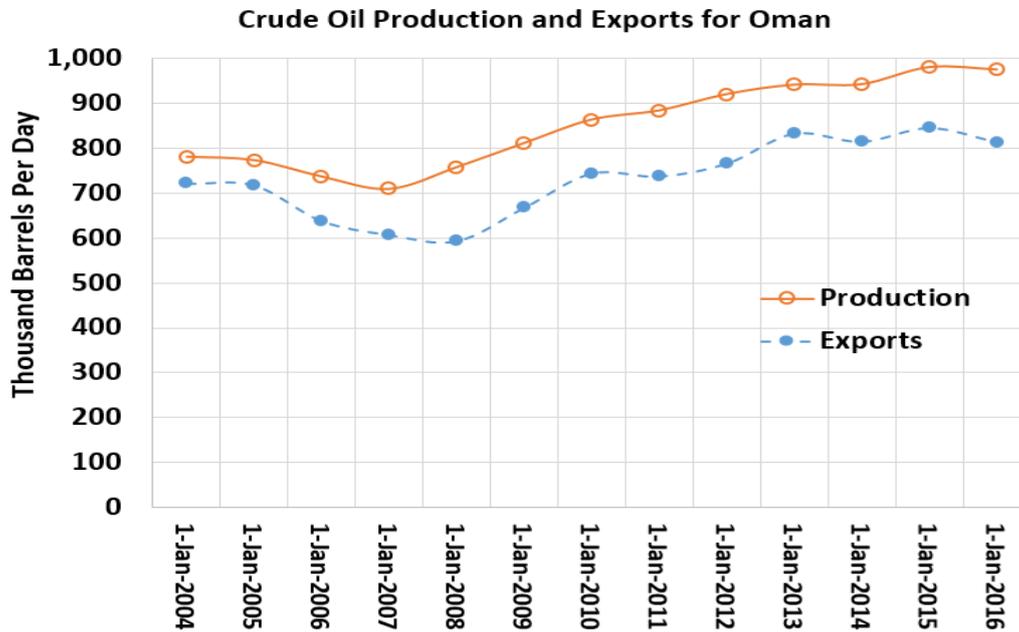

**Figure 8:** Historical data of the crude oil production and exports in Oman (in years). Data taken from[18].

## 3. Fall of Oil Prices in 2014/2015

In the second half of 2014, the oil prices fell down to about 40% of its high average price that it has been maintaining for more than three years. The crude oil Brent price per barrel dropped to 45.13 US$ on 13 January 2015. After a limited and unreliable recovery, a second phase of drop happened in the second half of 2015, and the barrel dropped further to 26.01 US$ on 20 January 2016. The price has recovered partially after this, but it is still far from what it was before 2014. Figure 9 visualizes these variations

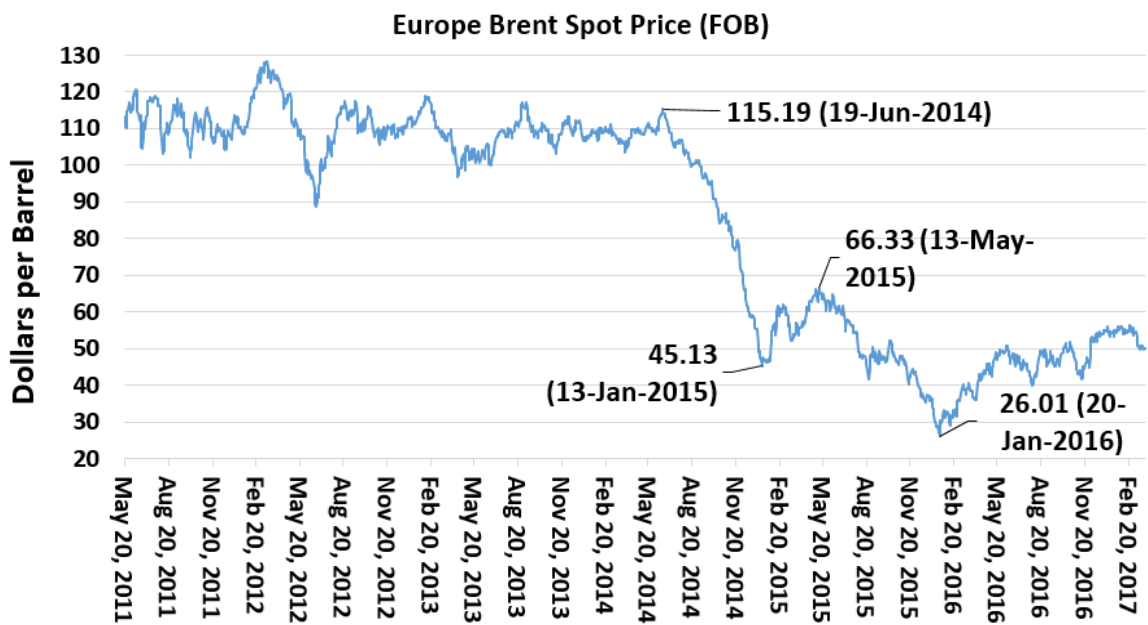

**Figure 9:** Historical data of the Brent spot prices (May2011–Feb2017). Data taken from [19].

Figure 10 gives a boarder look at the prices of Brent oil, and it shows that the 2014/2015 drop is not as strong as the earlier one of 2008. However, the price then did not reach the very low level as it did in the drop 2014/2015. In addition, the price went up relatively quickly after 2008, which was not the case in the 2014/2015 drop.

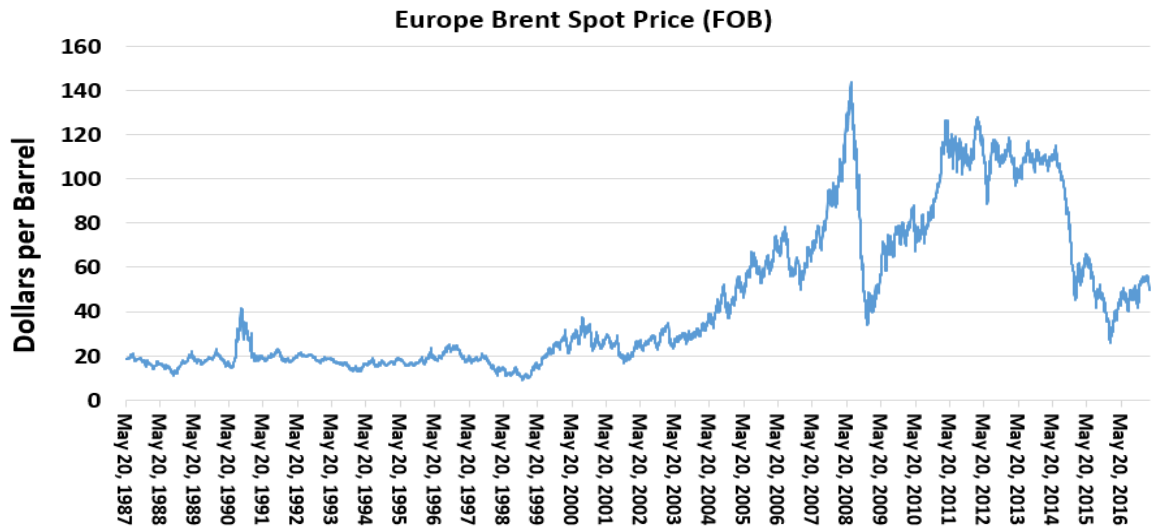

**Figure 10:** Historical data of the Brent spot prices (1987-2016). Data taken from[19].

Figure 11 presents the specific price variations for Oman oil. The percentage change (up or down) is indicated as vertical bars. The largest percentage drops (four) are highlighted with their numerical values. Out of these drops, the last two (2008 and 2014) are the strongest in terms of the absolute price change. In the same figure, the variations in the GDP per capita is plotted. The qualitative coherence between the GDP per capita and the oil price indicates the large dependence on oil, which is the main driver of the country's economy.

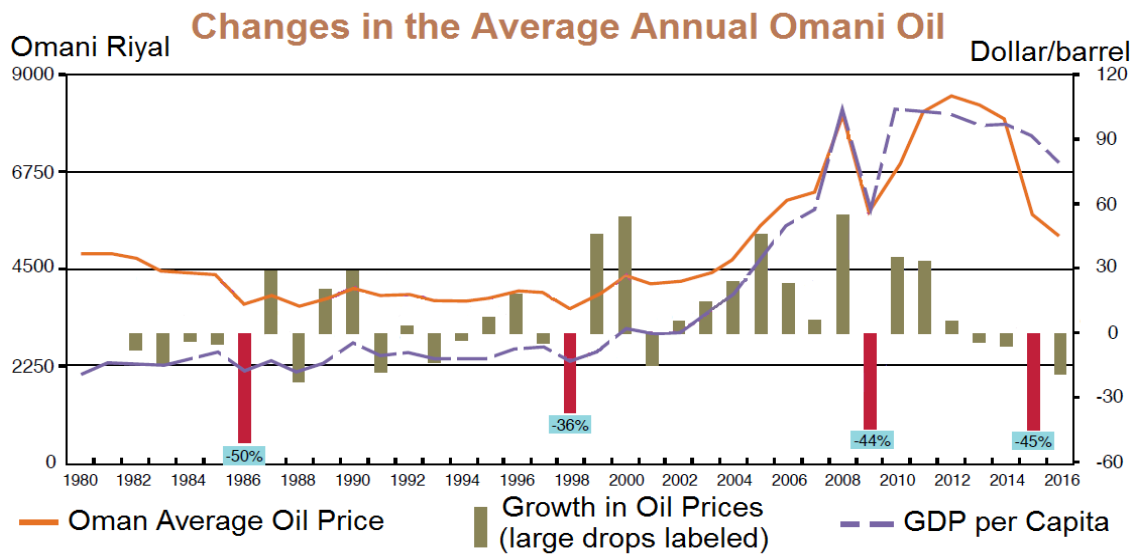

**Figure 11:** Historical data of the Omani oil price and its percentage change. The GDP per capita is also shown. Figure adapted from [20].

The recent decline in the oil price can be explained by four reasons as follows [21]:

  i.   The global demand for oil was low because of slowing economic activities, utilization of other fuels or energy sources, and increased energy efficiency.

  ii.  The U.S. imports of crude oil have decreased and the domestic production has increased rapidly after 2010, especially from the tight oil (shale oil), which is oil contained in low-permeable shale, sandstone, and carbonate rock formations [22]. These changes are demonstrated in Figures 12 and 13. The U.S. Energy Information Administration (EIA) made seven scenarios for the future of energy trade in the United States [23]. In five scenarios out of these seven, the U.S.A. is expected to become a net energy exporter (rather than importer). These projections are shown in Figure 14.

  iii. Saudi Arabia, Kuwait, Qatar, and United Arab Emirates decided not to curb production sharply to restore the oil price. This avoids sacrificing their market share, which could go to other countries such as Iran and Russia.

  iv.  Oil production in Iraq and Libya (two members of OPEC and large oil producers) continued despite their political instability. Their combined production is approximately 4 million barrels per day.

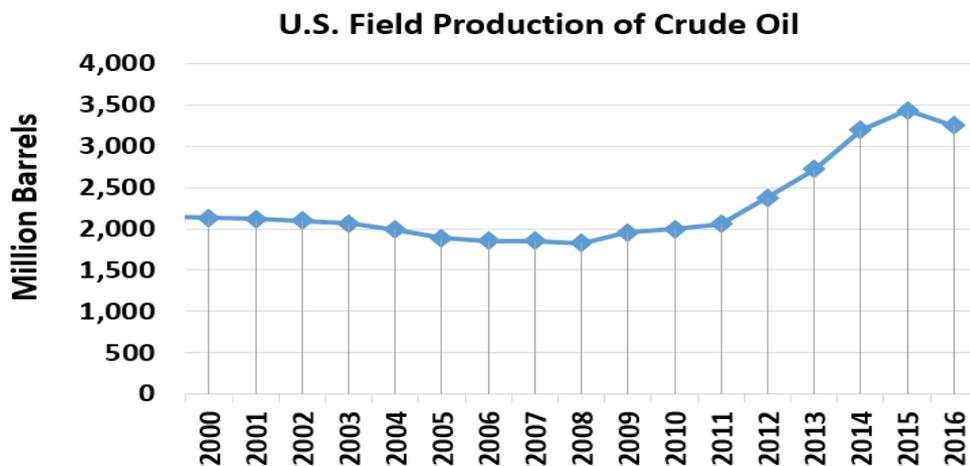

**Figure 12:** Historical data of the oil production in the U.S.A. Data from [24].

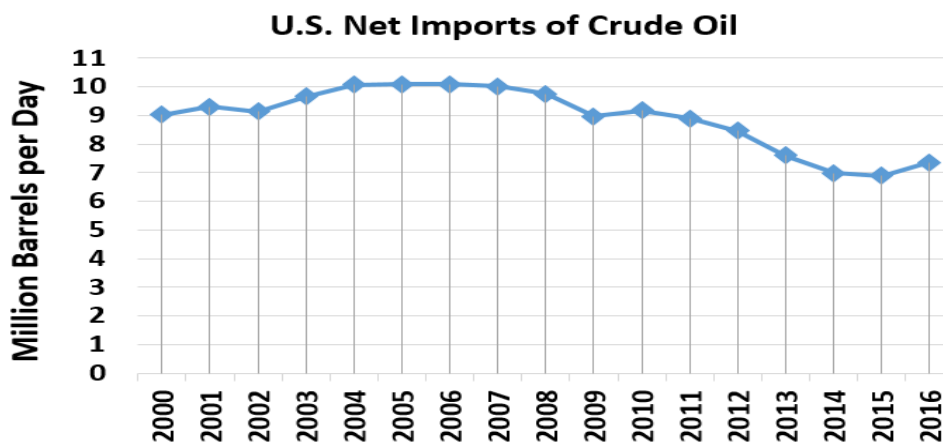

**Figure 13:** Historical data of the net imports of crude oil in the U.S.A. Data from [25].

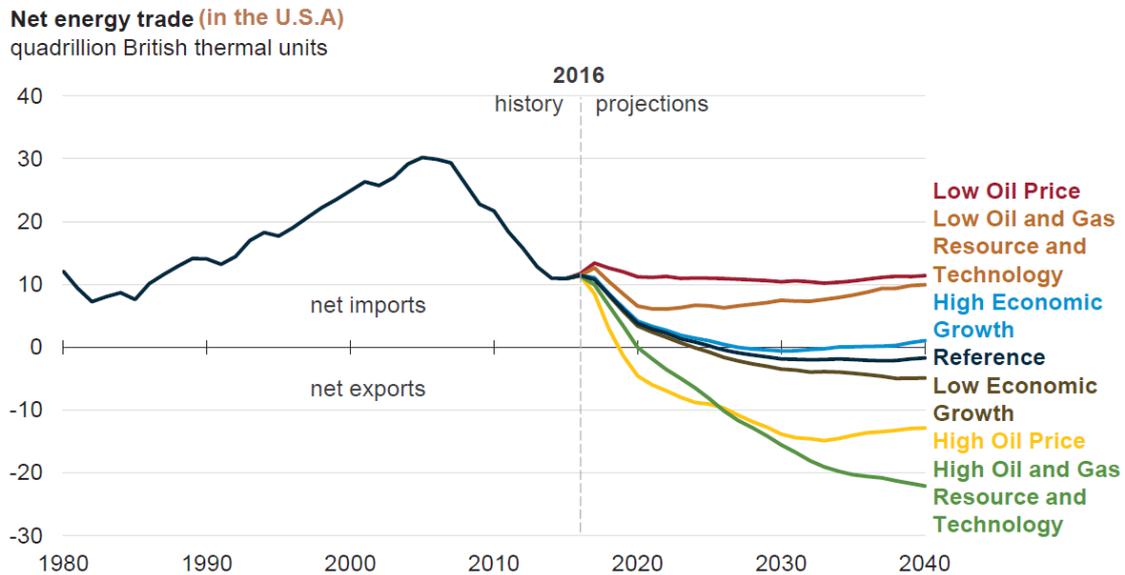

**Figure 14:** Historical and projected data for the net energy trade of the U.S.A. Figure adapted from [23].

## 4. Influences of the Oil Prices Fall in Oman

The recent decline in oil prices had several impacts on economy and public polices in Oman. As an oil-dependent country that has ongoing and long-term development plans (Oman Vision 2040 [26]), the decline has largely decreased the revenues, especially that the production of crude oil remained nearly flat, and mandated putting more stress on diversification of incomes [27]. Table 3 shows a summary of the annual budget of

Oman from 2014 to 2020 [20]. Oil prices are expected to recover partially over time, but they remain at about 50% of their value before the 2014 drop. It is worth mentioning that the 9[th] five-year plan for 2016-2020 included six different scenarios that vary based on the expected average price for the oil barrel over the plan's period, which ranged from 55 US$ to 75 US$. The data presented here correspond to the most probable scenario, in which the expected price is 55 US$ per barrel of oil.

**Table 3:** Key figures in the national budget for Oman.

| Year | 2014 | 2015 | 2016 | 2017 | 2018 | 2019 | 2020 |
|---|---|---|---|---|---|---|---|
| Average oil price, $ per barrel | 85 * | 75 * | 45 | 55 | 55 | 60 | 60 |
| Total revenues, million OMR | 11,700 | 11,600 | 8,600 | 9,800 | 10,300 | 11,100 | 11,300 |
| % of revenues from oil & gas | 82.5% | 79.0% | 71.5% | 73.1% | 71.1% | 71.8% | 71.3% |
| Tot. expenditure, million OMR | 13,500 | 14,100 | 11,900 | 12,700 | 13,300 | 13,900 | 14,100 |
| Deficit | 1,800 | 2,500 | 3,300 | 2,900 | 3,000 | 2,800 | 2,800 |
| % deficit to total revenues | 15% | 22% | 38% | 30% | 29% | 25% | 25% |
| * Actual value was 103.2$ for 2014[28]and 51.2$ for 2015[29] | | | | | | | |

In the following, we list some of the influences of the oil price drop in Oman:

i. **Growing fiscal deficit:**

Figure 15 shows that Oman has actually had a surplus in the national budget in 2013 (despite a provisional deficit). However, there has been an actual deficit in both 2014 and 2015. The provisional deficit is at its highest level in 2016 as was presented in Table 3.

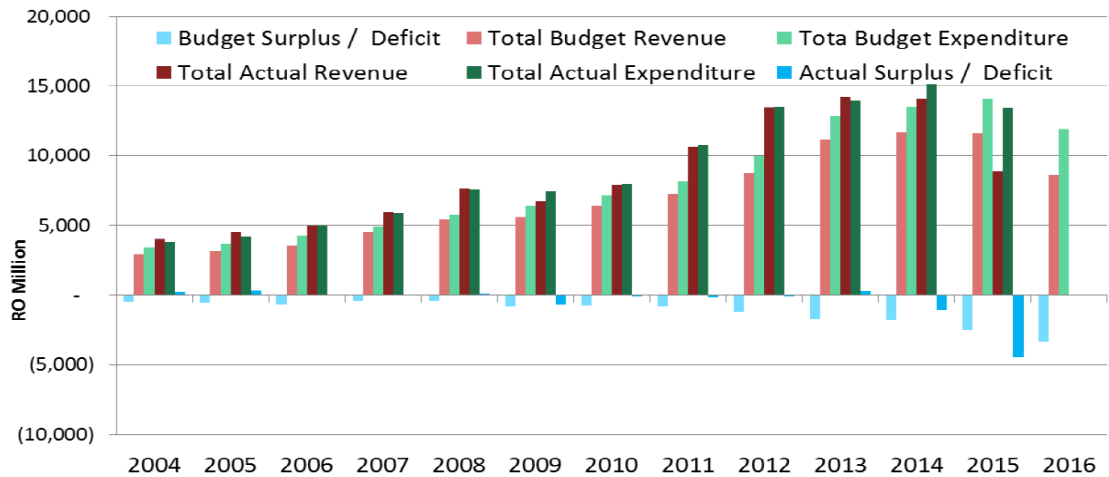

**Figure 15:** Historical data for the budgeted and actual revenue, expenditure, and deficit of Oman. Figure from [28].

ii. **Growing external debt:**

Oman used foreign borrowing as one of the means to finance the deficit. Figure 16 shows how the external debt (as a percentage of GDP) has increased rapidly after the 2014 drop in oil prices, where it jumped from 10.6% in the beginning of 2014 to 23% in the beginning of 2015, and then to about 30% in the beginning of 2016.

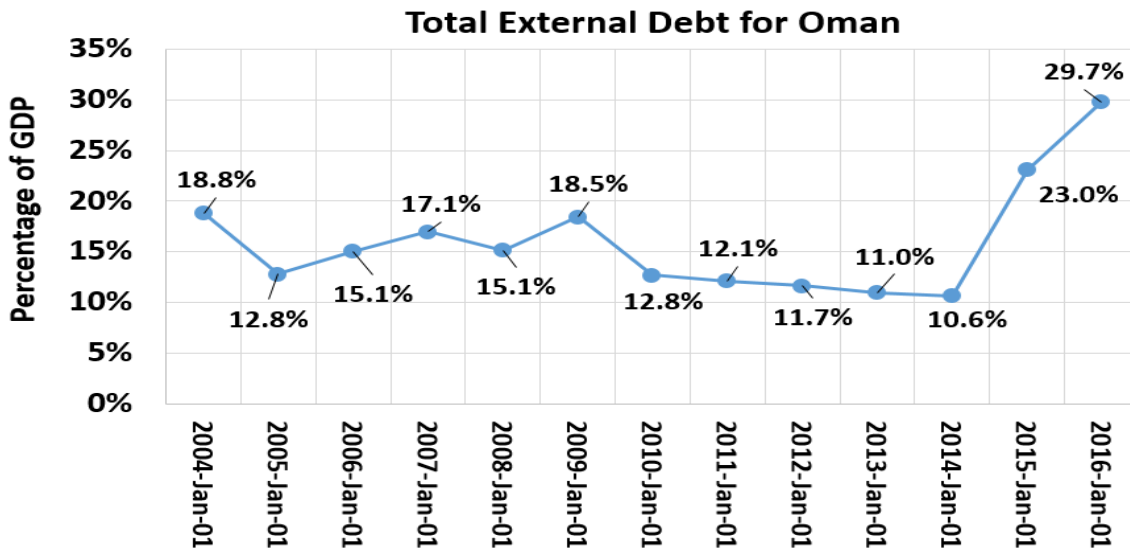

**Figure 16:** Historical data for the total external debt (as a percentage of the GDP) for Oman. Data from[30].

iii. **Declining national saving:**

Table 4 shows that the gross national saving (estimated) to the GDP for Oman decreased between 2014 and 2016 by a factor of about three [31]. The shown figures are calculated by deducting final consumption expenditure (household plus government) from the gross national disposable income. They contain the following items: personal saving, business saving (the sum of the capital consumption allowance and retained business profits), government saving (the excess of revenues over expenditures), but they exclude foreign saving (the excess of imports of goods and services over exports). The ranking of Oman in terms of national saving in 2016 was low, being the 154th.

**Table 4:** Recent values of the gross national saving (percentage of GDP) for Oman. Values are estimates.

| Year | 2014 | 2015 | 2016 |
|------|------|------|------|
| | 27.7% | 11.5% | 9.7% |

### iv.    Increased interest on deposits:

The Omani government used also other means to finance the fiscal deficit, which are domestic borrowing and withdrawing from governmental deposit/reserves. Such acts put some pressure on banks and caused them to increase the interest on deposits to attract more cash from the private sector. According to the Central Bank of Oman, the average interest on deposits (in local currency) increased from 0.910% in July 2015 to 1.219% in July 2016 [32].

### v.    Reduced subsidy:

The government in Oman and other five Gulf Cooperation Council (GCC) countries, in which Oman is a member, took measures to reduce expenditure by reducing subsidy of fuels [33]. This was aligned in Oman with an increase in the tariff of electricity [34] and water [35] (cutting down the subsidy).

For fuels, the rates were deregulated on 15 January 2016. The price of one liter of super unleaded gasoline (petrol) increased from 120 baisas[1] to 160 baisas. The liter of regular gasoline was pushed up from 114 baisas to 140 baisas. The liter of diesel fuel increased from 146 baisas to 160 baisas [36]. This fuel rise comes after about 17 years of steadiness, and it caused Oman's global ranking to go down from the 9th to the 13th among cheapest fuel providers. Fuel prices have increased further later, and are being adjusted on a monthly basis [37].

For electricity, the increase was limited to governmental, commercial, and industrial clients whose annual consumption exceeds 150 MWh. This band of heavy consumers accounts for only about 1% of the consumers, but their share in the overall consumption is about 30%. A new cost-reflective tariff (CRT) was announced on 12 October 2016 (effective 2017). In this new tariff plan, the concerned customers carry all expenses without any governmental subsidization. The new tariff plan has different rates over the day (dividing the day to a peak period and an off-peak period), over the week (dividing the week into weekdays and a weekend), and over the year (dividing the year into 5 intervals or seasons). The highest rate occurs during the summer season (May-July), in the weekday, at the peak period (from 1 pm to 5 pm). This highest rate is 0.061 OMR (0.16 US$) per kWh. The lowest rate occurs during the season of November-March, and it is rate (regardless of the day in the week or the hour in the day). That lowest rate is 0.012 OMR (0.3 US$) [38]. The residential sector has been excluded from the increase in the electricity tariff.

---

[1] 1 Omani Rial = 1,000 baisa

For water, the increase has also excluded the residential sector [39]. This of course reduces immediate burdens on the citizens and residents. The increases in the water tariff are summarized in Table 5.

**Table 5:** Changes in the water tariff in Omani baisas (US cents) per gallon*.

| Consumer class | Old rate | New rate |
|---|---|---|
| Government (up to 5,000 gal) | 2 (0.52) | 3.5 (0.91) Single tier |
| Government (above 5,000 gal) | 2.5 (0.65) | |
| commercial & industrial | 3 (0.78) | 3.5 (0.91) |
| Residential(up to 5,000 gal) | 2 (0.52) | SAME |
| Residential(above 5,000 gal) | 2.5 (0.65) | SAME |

* Oman uses the U.K. gallons, which is 4.55 liters

**vi.    Austerity measures:**
In its effort to cope with the shortfall in revenues (which is not expected to go away soon), the Omani government took a number of measures starting 2016 to reduce expenditures. One example of budget cuts include elimination of job benefits for employees of governmental authorities. These are more than ten authorities, and include the Public Authority for Civil Aviation (PACA), the Capital Market Authority's (CMA), the Public Authority for Electricity & Water (PAEW), and Telecommunications Regulatory Authority (TRA). These benefits include health insurance, life insurance, travel, car, loans, and cell (mobile) phones [40]. As another example, the Omani ministry of finance has issued a financial publication on 26 January 2016 with instructions against opening any tender or assigning any undertaking and accepting any financial commitment before referring to the ministry and seeking its approval. The ministry issued financial statement on 7 February 2016 regarding stopping external scholarships for studying abroad (toward bachelor, master, or doctoral degrees) at expense of governmental units [41].

**vii.    Economic diversification:**
Similar to neighbor oil-exporting countries, oil and gas in Oman are the dominant source of fiscal revenues [42]. In general, these countries should seek alternative sources of revenues keeping in mind that fossil fuels are exhaustible resources, and their prices are volatile. Also, the energy sector (which is typically highly capital intensive) creates few direct jobs. The recent fall of oil prices initiated focused attention to diversifying the country income. This matter is emphasized in the 9[th] five-year development plan of Oman (2016-2020) [20]. The plan's report points out that the previous 8th five-year plan (2011-2015) did not meet satisfactorily its aims with regard to the economic diversification, where oil and gas remained the predominant component in the revenues (with a share of 80%), and in the exporting income (with a share of 70%). The plan's report explains that the economic diversification has been relying on governmental spending and has been limited to non-tradable sectors (such as construction, transportation, and telecommunication). The prior efforts toward economic diversification have not targeted productive sectors that would increase exports. In the current plan, five potential diversification sectors have been identified as follows:

- Manufacturing industries
- Transportation and logistics
- Tourism
- Fisheries
- Mining

It is noted that, from Table 3, that the planned diversification is not expected to be dramatic since the portion of the revenues from oil and gas remains nearly at 71% during 2018-2020. Although this is an improvement over the value of 2014 (82.5%) and 2015 (79.0%), it is hoped that this figure can be pushed further down for more economic independence of oil price volatility.

viii.   **Facilitating investment:**
The public policies in Oman have paid attention to building an attractive environment for private domestic and foreign investments. This is reflected in the improvement in the position of Oman in the World Bank's 'Doing Business (DB)' ranking. This ranking is built on 10 indicators, which include starting a business, trading across borders, and enforcing contracts. The position of Oman went from 69[th] out of 189 (DB2016) to 66[th] out of 190 (DB2017) [43]. In particular, the gain is attributed largely to the indicator related to facilitating the starting of a business. This indicator is evaluated based on factors, such as the time and cost needed for: registration at the Commercial Registry of the Ministry of Commerce and Industry (MOCI), for registration at the Ministry of Manpower, for making a company seal, and for notification of the Tax Department of the Ministry of Finance. Oman's position for that particular indicator jumped from 159[th] in DB2016 to 32[nd] in DB2017.

## 5. Conclusions

This work considered the remarkable drop in oil prices in 2014 and then 2015, which affected both oil-exporting and oil-importing countries. The impact on oil-exporters is obviously adverse given the decline in revenue. The work examined the Sultanate of Oman (Oman) as a typical oil-exporting country, which still relies dominantly on oil and gas for revenues and exporting. The country faced challenges due to the shrinkage in its GDP and revenue, which occur during a period of ambitious development and national transformation. These issues were discussed, and then some measures taken by the country to respond safely to them were presented with some specific examples. The worst (bottleneck) fiscal year in terms of the budgeted deficit was 2016. It is worth mentioning here that the exchange rate for the Omani rial and US$ did not change, and this remains a primary policy objective for the Central Bank of Oman [6]. In addition, the annual touristic festival "Muscat 2016" was held (14 January to 13 February 2016) as regular and was not canceled [44]. These are indications that the country could safely protect its economy from the instability it faced.

## Bibliography

1. Food and Agriculture Organization of the United Nations (FAO), "Oman's Profile," AQUASTAT website 2016, 2016. [Online]. Available: http://www.fao.org/nr/water/aquastat/countries_regions/omn/. [Accessed 01 04 2017].

2. "Physical Location Map of Oman, highlighted continent," Maphill, [Online]. Available: http://www.maphill.com/oman/location-maps/physical-map/highlighted-continent/. [Accessed 01 04 2017].

3. "The World Bank Open Data, Total Pouplation of Oman," The World Bank Group, [Online]. Available: http://data.worldbank.org/indicator/SP.POP.TOTL?locations=OM. [Accessed 01 04 2017].

4. Times News Service, "Number of expats in Oman reaches 2 million," Times of Oman, 19 April 2016.

5. The World Bank Data, "Population growth (annual %) of Oman," The World Bank Group, [Online]. Available: http://data.worldbank.org/indicator/SP.POP.GROW? locations=OM. [Accessed 01 04 2017].

6. H.E. Hamood bin Sangour Al Zadjali as an Overall Supervisor (Executive President of the Central Bank of Oman), "Financial Stability in Oman," Al Markazi, a bi-monthly publication of the Central Bank of Oman, p. 30, October 2016.

7. K. Alzadjali, "46% of Oman's foreign investment is from UK," Times of Oman, 22 March 2017.

8. Vision of Humanity, "Global Peace Index," Institute for Economics and Peace (IEP), [Online]. Available: http://static.visionofhumanity.org/#/page/indexes/global-peace-index. [Accessed 23 3 2017].

9. M. D. Petraglia and J. I. Rose, The Evolution of Human Populations in Arabia: Paleoenvironments, Prehistory and Genetics, Springer Science & Business Media, 2009.

10. A. Hamilton, An Arabian Utopia: The Western Discovery of Oman, Arcadian Library, 2010.

11. AQUASTAT (Global Water Information System), "Computation of long-term annual renewable water resources (RWR) by country - Oman," FAO (Food and Agriculture Organization of the United Nations), 2016.

12. World Bank Data, "GDP (current US$)," The World Bank Group, [Online]. Available: http://data.worldbank.org/indicator/NY.GDP.MKTP.CD?locations=OM. [Accessed 01 04 2017].

13. World Bank Data, "GNI per capita, Atlas method (current US$)," The World Bank Group, [Online]. Available: http://data.worldbank.org/indicator/NY.GNP.PCAP.CD?locations=OM&view=chart. [Accessed 01 04 2017].

14. "World Population Prospects: Key findings & advance tables," United Nations, Department of Economic and Social Affairs, Population Division, New York, 2015.

15. World Bank Data, "Life expectancy at birth, total (years)," The World Bank Group, [Online]. Available: http://data.worldbank.org/indicator/SP.DYN.LE00.IN. [Accessed 01 04 2017].

16. Selim Jahan (Director and lead author), "Human Development Report 2016 Human Development for Everyone," The United Nations Development Programme (UNDP), New York, 2016.

17. "World Urbanization Prospects: The 2014 Revision," United Nations, Department of Economic and Social Affairs, Population Division. [Online]. [Accessed 02 04 2017].

18. "FRED® Economic Data (Oman)," Federal Reserve Bank of St. Louis, [Online]. Available: https://fred.stlouisfed.org/categories/32780. [Accessed 02 04 2017].

19. Spot Prices for Crude Oil and Petroleum Products, "Europe Brent Spot Price FOB (Dollars per Barrel)," The U.S. Energy Information Administration (EIA), [Online]. Available: https://www.eia.gov/dnav/pet/hist/LeafHandler.ashx?n=pet&s=rbrte&f=d. [Accessed 03 04 2017].

20. Sultanate of Oman Supreme Council for Planning, "Summary of the Ninth Five-Year Development Plan (2016 - 2020)," Muscat, Oman, 2016.

21. E. L., "Why the oil price is falling," The Economist Newspaper, 8 12 2014. [Online]. Available: http://www.economist.com/blogs/economist-explains/2014/12/economist-explains-4. [Accessed 03 04 2017].

22. "How much shale (tight) oil is produced in the United States?," U.S. Energy Information Administration, 13 2 2017. [Online]. Available: https://www.eia.gov/tools/faqs/faq.php?id=847&t=6. [Accessed 3 4 2017].

23. "Annual Energy Outlook 2017 (with projections to 2050)," U.S. Energy Information Administration, Washington, D.C., United States, 2017.

24. "U.S. Field Production of Crude Oil," U.S. Energy Information Administration, [Online]. Available: https://www.eia.gov/dnav/pet/hist/LeafHandler.ashx?n=pet&s=mcrfpus1&f=a. [Accessed 03 04 2017].

25. "U.S. Net Imports of Crude Oil," U.S. Energy Information Administration (EIA), [Online]. Available: https://www.eia.gov/dnav/pet/hist/LeafHandler.ashx?n=PET&s=MCRNTUS2&f=A. [Accessed 03 04 2017].

26. P. Joyce and T. F. Al Rasheed, Public Governance and Strategic Management Capabilities: Public Governance in the Gulf State, Routledge, 2016.

27. S. Hasan, "Oman could overcome fiscal challenges with diversification: Moody's report," 25 3 2017. [Online]. Available: http://timesofoman.com/article/105652/Business/Oman-could-overcome-fiscal-challenges-withdiversification:-Moody's-report. [Accessed 4 4 2017].

28. "Annual Report 2014," Central Bank of Oman, Muscat, 2015.

29. Investment Banking Division, Economic Research, "An Insight into Oman's State Budget 2016," National Bank of Oman, Muscat, 2016.

30. "Total External Debt for Oman," Federal Reserve Bank of St. Louis, 20 1 2017. [Online]. Available: https://fred.stlouisfed.org/series/OMNDGDPGDPPT. [Accessed 4 4 2017].

31. [31]   31.   U.S. Central Intelligence Agency (CIA), "The World Factbook (Oman)," [Online]. Available: https://www.cia.gov/library/publications/the-world-factbook/geos/mu.html. [Accessed 4 4 2017].

32. "Increased avearge interest in Oman on loans and deposits (in Arabic)," Al-araby Al-gadeed, Muscat, 2016.

33. Aliqtisadi, "Despite lifting the subsidy .. the fuel in the Gulf remains the cheapest internationally (in Arabic)," 7 5 2016. [Online]. Available: https://ae.aliqtisadi.com/778224-%D8%A3%D8%B3%D8%B9%D8%A7%D8%B1-%D8%A7%D9%84%D9%88%D9%82%D9%88%D8%AF-%D9%81%D9%8A-%D8%AF%D9%88%D9%84-%D8%A7%D9%84%D8%AE%D9%84%D9%8A%D8%AC/. [Accessed 8 4 2017].

34. "The Sultanate of Oman raises the electricity tariff starting 2017 (in Arabic)," Al-araby Al-gadeed, Muscat, 2016.

35. "Adjusting the tariff of electricity and water (in Arabic)," Atheer, Muscat, 2016.

36. R. K, "Oman hikes fuel prices by 50 per cent in five months," Times of Oman, 31 5 2016. [Online]. Available: http://timesofoman.com/article/85125/Oman/Government/Oman-hikes-fuel-prices-by-50-per-cent-in-five-months. [Accessed 6 4 2017].

37. R. K, "Petrol price hike: Oman to fall from ninth cheapest fuel in world to 13[th] Times of Oman," Times of Oman, 11 1 2016. [Online]. Available: http://timesofoman.com/article/75319/Oman/Transport/Price-of-fuel-prices-set-to-rise-in-Oman-from-January-15. [Accessed 6 4 2017].

38. Muscat Electricity Distribution Company (member of Nama Group), "The cost-reflective tariff, CRT. The details according to the rates of 2017 (in Arabic)," [Online]. Available: http://www.medcoman.com/PDF/crt-ar-2017.pdf. [Accessed 1 4 2017].

39. Oman Public Authority for Electricity & Water (PAEW), "A new tariff for water approved by the government starting from the bills of March for government, industrial, and business consumption (in Arabic)," 3 2016. [Online]. Available: https://www.paew.gov.om/Library/News/New-water-tariff-announced-for-government,-commerc?lang=ar-OM. [Accessed 12 3 2017].

40. "Austeristy is launched in the Sultanate of Oman (in Arabic)," mangish.net, 1 5 2016. [Online]. Available: http://mangish.net/%D8%A7%D9%84%D8%AA%D9%82%D8%B4%D9%81-%D9%8A%D9%86%D8%B7%D9%84%D9%82-%D9%81%D9%8A-%D8%B3%D9%84%D8%B7%D9%86%D8%A9-%D8%B9%D9%85%D8%A7%D9%86/. [Accessed 6 4 2017].

41. "The Sultanate of Oman declares austerity (in Arabic)," http://nononline.net, 26 3 2016. [Online]. Available:

http://nononline.net/%D8%B3%D9%84%D8%B7%D9%86%D8%A9-%D8%B9%D9%8F%D9%85%D8%A7%D9%86-%D8%AA%D8%B9%D9%84%D9%86-%D8%A7%D9%84%D8%AA%D9%82D8%B4%D9%81/. [Accessed 6 4 2017].

42. International Monetary Fund, "Economic Diversification in Oil-Exporting Arab Countries," Annual Meeting of Arab Ministers of Finance, Manama, Bahrain, 2016.

43. International Bank for Reconstruction and Development, "Doing Business 2017 - Economy Profile 2017 Oman," The World Bank, Washington DC, 2017.

44. Muscat Daily, "Muscat Festival to be held from Jan 14-Feb 13, says organising committee," 5 10 2015. [Online]. Available: http://www.muscatdaily.com/Archive/Oman/Muscat-Festival-to-be-held-from-Jan-14-Feb-13-says-organising-committee-4c91. [Accessed 8 4 2017].